\documentclass[Letter]{aa}

\usepackage{caption}
\usepackage{subcaption}
\usepackage{lipsum}
\usepackage{lscape}
\usepackage{amsmath}
\usepackage{xcolor}
\usepackage{booktabs}
\usepackage[symbol]{footmisc}
\usepackage{orcidlink} 
\usepackage{comment}
\usepackage{lastpage}

\usepackage{hyperref}
\hypersetup{
  colorlinks   = true, 
  urlcolor     = blue, 
  linkcolor    = blue, 
  citecolor   = blue 
}

\usepackage{graphicx}
\usepackage{txfonts}
\begin{document}

   \title{A redshift-independent theoretical halo mass function validated with \textsc{Uchuu} simulations}

   \author{Elena Fernández-García\orcidlink{0009-0006-2125-9590}\thanks{e-mail: efdez$@$iaa.es}
          \inst{1} 
          \and
          Juan E. Betancort-Rijo
          \inst{2, 3}
          \and
          Francisco Prada\orcidlink{0000-0001-7145-8674}
          \inst{1,2}
          \and
          Tomoaki Ishiyama\orcidlink{0000-0002-5316-9171}
          \inst{4}
          \and
          Anatoly Klypin
          \inst{5}
          \and
          José Ruedas
          \inst{1}
          }

   \institute{Instituto de Astrofisica de Andalucia (CSIC), E18008 Granada, Spain,
         \and
             Instituto de Astrofisica de Canarias, C/ Via Lactea s/n, Tenerife E38200, Spain
            \and 
            Facultad de Fisica, Universidad de La Laguna, Astrofisico Francisco Sanchez, s/n, La Laguna, Tenerife E38200, Spain
            \and 
            Digital Transformation Enhancement Council, Chiba University, 1-33, Yayoi-cho, Inage-ku, Chiba, 263-8522, Japan
            \and
            Department of Astronomy, University of Virginia, Charlottesville, VA 22904, USA
             }

   \date{Received ..., ...; accepted ..., ...}


    \abstract
    {We present a new theoretical framework for the halo mass function (HMF) that accurately predicts the abundance of dark matter halos over an exceptionally wide range of masses and redshifts, based on a generalised Press–Schechter model with triaxial collapse (GPS+). The HMF is formulated mainly as a function of the variance of the linear density field, with a weak explicit mass dependence and no explicit redshift dependence, which is able to naturally reproduce the correct normalisation and high-mass behaviour without requiring an empirical fitting. Using the \textsc{Uchuu} $N$-body simulation suite under \textit{Planck} cosmology, combining six simulations with up to 300 realisations, we measured the HMF over $6.5 \leq \log(M_{200m}/[h^{-1}M_\odot]) \leq 16$ and $0 \leq z \leq 20$ with reduced cosmic variance. Over this full domain, we find that GPS+ matches the simulations to within $10$–$20\%$, performing similarly to the Sheth–Tormen model at $z \lesssim 2,$ but with substantially results at higher redshifts. In the latter case, the Sheth–Tormen model can deviate by $70$–$80\%,$ while GPS+ will remain within $\sim20\%$. Finally, we show that the halo mass definition is key: $M_{200m}$ yields a nearly universal, weakly redshift-dependent HMF, whereas adopting the evolving virial overdensity from Bryan $\&$ Norman (1998) ends up degrading the agreement at low redshifts and high masses.
    }


   \maketitle
%

\section{Introduction}

The use of the halo mass function (HMF) to describe the abundance of dark matter halos as a function of mass is a cornerstone of modern cosmology. It encodes the growth of structure from the early Universe to the present day. It provides a fundamental link between theory, simulations, and observations, making it a key probe of cosmological models and parameters \citep{2002PhR...372....1C, 2010gfe..book.....M, 2016MNRAS.463.1666C}. Accurate predictions of its mass and redshift dependence are therefore essential for interpreting data from current and upcoming large-scale structure surveys.

The theoretical foundation of the HMF was laid by the Press--Schechter (PS) formalism \citep{1974ApJ...187..425P}, later developed through excursion-set theory \citep{1991ApJ...379..440B, 1993MNRAS.262..627L} and refined with ellipsoidal-collapse models \cite[][hereafter ST]{1999MNRAS.308..119S, 2001MNRAS.323....1S}. Empirical calibrations based on $N$-body simulations have since improved its accuracy over a range of cosmologies, halo masses, and redshifts \citep[e.g.][]{2001MNRAS.321..372J, 2008ApJ...688..709T, 2016MNRAS.456.2486D, Shirasaki2021, 2021A&A...652A.155S}. However, all existing fits remain limited in scope: early models can be applied mainly to cluster-scale halos at $z\lesssim3$, while later extensions reach $z\lesssim10$, but typically only for halo masses above $10^{10}\,h^{-1}M_\odot$. As a result, no current model provides a self-consistent description of the HMF across the full halo mass hierarchy and cosmic history.

Recent advances in numerical simulations, particularly the \textsc{Uchuu} $N$-body simulation suite (\citealt{2021MNRAS.506.4210I}; Ishiyama et al., in prep.), make it possible to overcome many of these limitations by combining extremely large cosmological volumes with a high mass resolution. This enables precise measurements of the HMF over a vastly extended range of halo masses and redshifts, offering a unique opportunity to revisit and improve theoretical models.

In this work, we build on the generalised Press-Schechter formalism of \citet{2006ApJ...650L..95B, 2006ApJ...653L..77B} and present a new model, named \emph{GPS+}, which delivers a consistent description of the HMF from dwarf-scale halos to the most massive clusters, spanning $6.5<\log(M/M_\odot)<16$ and redshifts $0\le z\le20$. While it remains rooted in the generalised PS framework, triaxial collapse plays a central role in the construction of the model. With minimal calibration to the \textsc{Uchuu} simulations, the resulting HMF depends primarily on the variance of the linear density field, $\sigma(M,z)$, exhibiting only weak explicit mass dependence and with no explicit dependence on redshift. This unprecedented dynamical range and per cent-level accuracy are enabled by the scope and quality of the \textsc{Uchuu} suite, providing a robust framework for interpreting data from ongoing and future surveys such as DESI, Euclid, Subaru~PFS, LSST, eROSITA, and JWST.

While our approach is still closely connected to the classical PS picture, it replaces the original collapse assignment with a more physically motivated prescription, while preserving the statistical properties of Gaussian random fields and the conceptual structure of Press-Schechter theory.

This paper is organised as follows. Section~\ref{simulations} describes the \textsc{Uchuu} simulations and HMF measurements. The GPS+ theoretical framework is presented in Section~\ref{theoretical_framework}. In Section~\ref{results}, we compare the model predictions with simulation results and Section~\ref{summary} summarises our conclusions.

\section{Uchuu simulations}\label{simulations}
In this work, we tested the theoretical framework using the \textsc{Uchuu} simulation suite spanning a wide range of mass resolutions, including \textsc{Phi-4096}, \textsc{Shin-Uchuu}, \textsc{Uchuu}, \textsc{Mucho-Uchuu-140M}, \textsc{Mucho-Uchuu-1G}, and \textsc{Mucho-Uchuu-6G}\footnote{All simulations are publicly available at \url{https://skiesanduniverses.org/Simulations/Uchuu/}}, as described in \citealt{2021MNRAS.506.4210I}; Ishiyama et al., in prep. Their main properties are summarised in Table~\ref{sims}. The first three simulations are based on  \textit{Planck15} cosmology, while the remaining three follow \textit{Planck18}. Despite these slight cosmological differences, their impact on the HMF is negligible compared to both the intrinsic statistical uncertainties of the simulations and the theoretical modelling uncertainties.

We used 50, 300, and 100 independent realisations of \textsc{Mucho-Uchuu-140M}, \textsc{Mucho-Uchuu-1G}, and \textsc{Mucho-Uchuu-6G}, respectively. The HMF is obtained by averaging over all realisations, with uncertainties estimated from the diagonal of the covariance matrix, an ensemble approach that significantly reduces cosmic variance. For the purposes of illustration, within the total volume of $21,600\,h^{-3}\,\mathrm{Gpc}^3$ covered by the 100 \textsc{Mucho-Uchuu-6G} realisations, the most massive halo is characterised by $M = 8.3 \times 10^{15},M_\odot$, about 13 times the mass of the Virgo cluster \citep{2020A&A...635A.135K}.

All simulations were run with the \textit{TreePM} code \textit{GreeM} \citep{art3, art4}. \textsc{Phi-4096}, \textsc{Shin-Uchuu}, and \textsc{Uchuu} were executed on the ATERUI II supercomputer at CfCA, National Astronomical Observatory of Japan, while the \textsc{Mucho-Uchuu} simulations used the FUGAKU supercomputer at RIKEN, Kobe. Halos were identified with the \textit{RockStar} halo/subhalo finder \citep{Behroozi2013} and its distributed version \textit{MPI-Rockstar}\footnote[3]{\url{https://github.com/Tomoaki-Ishiyama/mpi-rockstar/}} \citep{Tokuue2024}, and merger trees were constructed using \textit{Consistent Trees} \citep{Behroozi2013b}.

\begin{figure}
    \centering
    \includegraphics[width=1.0\linewidth]{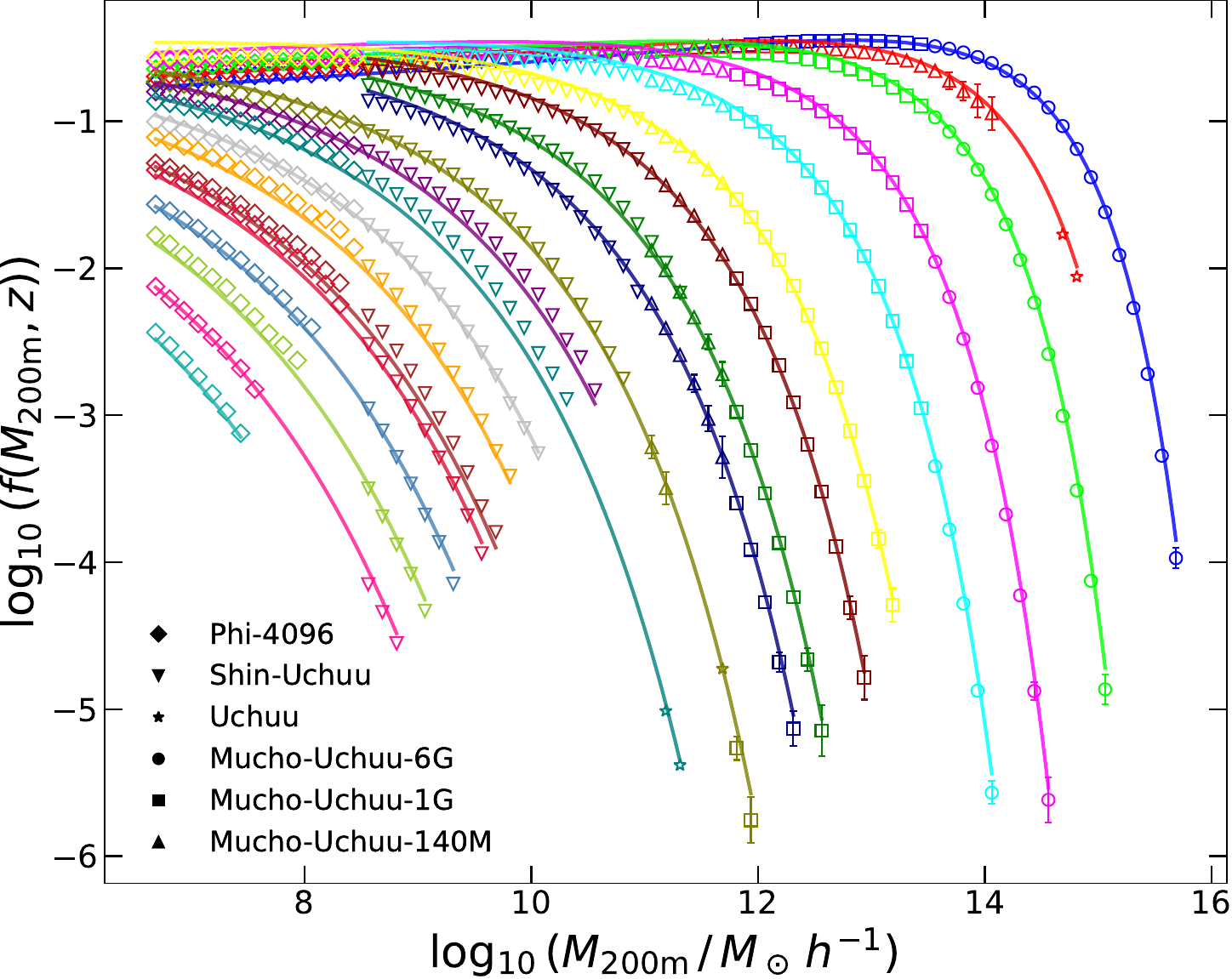}
    \caption{Multiplicity function $f(M,z)$ (Eq.~\ref{multfunc}) from \textsc{Uchuu} simulations (markers) vs GPS+ predictions (solid lines) at $z=0$–20 (right to left; see Table~\ref{redshifts}). The $y$-axis shows the normalised differential halo abundance including $d\ln M/d\ln \sigma$. Only bins with $>200$ halos are shown.}

    \label{hmf_data_plot}
\end{figure}

Figure \ref{hmf_data_plot} shows the combined HMF derived from all the simulations used in this work. We express it in terms of the the multiplicity function \cite[e.g][]{1999MNRAS.308..119S}, defined as

\begin{equation}
    f(M, z) =  
\dfrac{M_{\rm 200m}}{\rho_{\rm m}}
\dfrac{dn}{d\ln M_{\rm 200m}}
\left| \dfrac{d\ln M_{\rm 200m}}{d\ln \sigma} \right|.
\label{multfunc}
\end{equation}

We adopted the halo mass M$_{\rm 200m}$, defined within an overdensity of 200$\rho_{\rm b}$, where $\rho_{\rm b}$ is the mean background density. Unlike the virial mass, this fixed-threshold definition is redshift-independent. We refer to Appendix~\ref{mass_def} for more details.

To construct the combined HMF at each redshift, we used halos with at least 407 particles to ensure low-mass completeness. The high-mass end of each simulation is limited to allow for a smooth, continuous transition across overlapping mass ranges. The resulting HMFs are available at the CDS (see Appendix~\ref{hmf_data}).

\section{Theoretical framework}\label{theoretical_framework}

We modelled the HMF using the GPS+ framework, which describes halo abundances over an exceptionally wide mass range ($6.5 \lesssim \log[M/(h^{-1}M_\odot)] \lesssim 16$) and up to very high redshifts ($z \sim 20$). The formalism builds on the PS idea that a mass element belongs to a halo of mass larger than $M$ if the smoothed density field satisfies a collapse condition; however, it replaces the original centre-based criterion with a more physically motivated prescription. The model preserves the Gaussian-field statistics underlying PS and excursion-set theory while incorporating the physics of triaxial collapse. In this way, GPS+ retains the conceptual structure of PS but modifies the criterion for assigning mass elements to collapsed objects of mass larger than $M$.

In this framework, the HMF is given by \citep{MonteroDorta2006}:

\begin{equation}
    \frac{dn(M,z)}{dM} = \frac{\rho_m}{M} \frac{dF(M,z)}{dM},
    \label{hmf_eq}
\end{equation}
where $\rho_{\rm m}$ is the matter density, and $F(M,z)$ is the mass fraction.

The mass fraction is defined as

\begin{equation}
    F(M) = \frac{ {\rm erfc} \left[ <\delta_c(\sigma(M,z), M)> / (\sqrt{2} \, \sigma(M,z)) \right] }{ V(\Sigma(M,z), M) },
    \label{Fm}
\end{equation}
with $V(\Sigma(M,z), M)$ given by

\begin{equation}
\begin{split}
    V(\Sigma(M,z), M) = 3 \int_0^1 
    {\rm erfc}\Bigg[ \frac{\langle \delta_c \rangle(\sigma(M,z), M)}{\sqrt{2}\,\Sigma(M,z)}\times
    & \\
    \times \sqrt{\frac{1 - \exp(-c(M)\,\xi^2)}{1 + \exp(-c(M)\,\xi^2)}} 
    \Bigg]\, \xi^2 \, d\xi
\end{split}
\label{Vm}
,\end{equation}
where $\delta_l$ encodes the linear growth corrections at $z=0$, $\Sigma(M,z)$ is given by
\begin{equation}
    \Sigma(M,z) = \sqrt{\sigma^2(M,z) + U^2(\sigma/\delta_c)},
\end{equation}
where $\sigma(M,z)$ is the standard linear-theory variance, and the correction function $U(x)$ is defined as

\begin{equation}
    U(x) = -0.01507 + 0.17810\cdot\, x + 0.03835\cdot\, x^2 -0.00221\cdot\, x^3,
\label{Ux}
\end{equation}
where $x = \sigma(M,z)/\delta_{c}$, and $\delta_{c}$ is the critical overdensity, equal to the linear-theory density contrast, $\delta_{l}$, in the spherical-collapse approximation at collapse. In our model, the critical overdensity for collapse is further modified as a function of mass and redshift, expressed as

\begin{equation}
\begin{aligned}
\langle \delta_c \rangle
&= \delta_{c}
\left(1 + 0.845\,x - 0.04\,x^2 + 0.0025\,x^3\right)^{B}\times \\
&\quad \times A
\left(1 + 0.17\,b(M) - 0.087\,b^2(M)\right)^{D}
\end{aligned}
\label{delta_c}
\end{equation}

where $b(M)$ is defined by
\begin{equation}
\begin{split}
\log b(x) &= -1.28 + 0.05781\,x - 0.005622\,x^{2} \\
          &\quad + 5.884\times10^{-4}\,x^{3}
          - 1.365\times10^{-5}\,x^{4},
\end{split}
\label{bm}
\end{equation}

where $x=\log{(M/[M_{\odot}/h])}$, while $A$ and $B$ are free parameters fitted to the simulation data (see values below) and $D$ is a parameter whose theoretical value is 1 (see Betancort-Rijo et al., in prep.). However, we did allow it to to test the robustness of this prediction. Equation \ref{delta_c} essentially comes  from the physics of the triaxial collapse (see Appendix \ref{formalism_eqs} for the details).

The function $c(M)$ from Equation \ref{Vm} is a polynomial  defined as

\begin{equation}
\begin{split}
    \log c(x) &= - 1.124+ 0.01756\cdot x + 0.002539\cdot x^{2} - \\
    &\quad - 6.438\times10^{-5}\cdot x^{3} +4.726\times10^{-6}\cdot x^{4},
\end{split}
\label{cm}
\end{equation}
where $x=\log{(M/[M_{\odot}/h])}$. The quantities entering Eqs. \ref{Ux}–\ref{delta_c} follow from the physics of triaxial collapse: $U^2$ can be obtained from first-principles calculations in \cite{MonteroDorta2006} and Eq. \ref{Ux} is a fit to those results. The functions $b(M)$ and $c(M)$ are fully determined by the linear matter power spectrum \citep{2007MNRAS.378..339S,2006ApJ...653L..77B}. The polynomials given by Eqs. \ref{bm} and \ref{cm} correspond to fits done in this work for the Planck power spectrum.

This formulation enables the computation of the HMF across a broad halo mass range and out to very high redshifts. A concise overview of the theoretical framework is provided in Appendix~\ref{formalism_eqs}, while a full derivation of the equations will be presented in Betancort-Rijo et al. in prep. 

The coefficients \(A\) and \(B\) in Eq.~\ref{delta_c} were obtained by fitting Eq.~\ref{hmf_eq} to the simulation data using the statistic,
\begin{equation}
    \mho^{2} = \sum_{i,j}\frac{|n_{\rm sim}(M_{i},z_{j})-n_{\rm th}(M_{i},z_{j})|}{n_{\rm sim}(M_{i},z_{j})^{2}}
    \left[1-\Theta\!\left(\frac{\Delta n_{\rm sim}(M_{i},z_{j})}{0.05\,n_{\rm sim}(M_{i},z_{j})}-1\right)\right],
\end{equation}
where \(n_{\rm sim}\) and \(n_{\rm th}\) are the simulated and theoretical HMFs, and \(\Delta n_{\rm sim}\) is the statistical uncertainty. The Heaviside function \(\Theta\) restricts the fit to bins with relative uncertainties below \(5\%\).

Uncertainties were assumed to be Poissonian for simulations with a single realisation (\textsc{Phi-4096}, \textsc{Shin-Uchuu}, \textsc{Uchuu}), while for \textsc{Mucho-Uchuu-140M}, \textsc{Mucho-Uchuu-1G}, and \textsc{Mucho-Uchuu-6G,} they were derived from the diagonal of the covariance matrix over multiple realisations.

This fit gives $A=1.089$ and $B=0.652$. Varying $D$ indicates the best agreement at $D\simeq 1$, consistent with theory. We note that only $A$ and $B$ were fitted to the simulations and their values are close to the expected theoretical value of 1.

A key feature of this formalism is that \(f(\sigma)\) depends explicitly on mass through the functions \(b(m)\) and \(c(m)\) (Eqs.~\ref{bm} and~\ref{cm}), with no explicit dependence on redshift. If this mass dependence is (incorrectly) neglected and we assume that \(f\) depends only on \(\sigma\), we then obtain $f(m,z)=f(\sigma(m,z),m)=f(\sigma,m(\sigma,z))=f(\sigma,z)$. The first equality is correct: at fixed mass, any redshift dependence enters only through \(\sigma(m,z)\). The second equality instead removes the explicit mass dependence of \(f\) at fixed \(\sigma\) (a central aspect of our framework) by using the \(m\)--\(\sigma\) relation at fixed redshift, thereby introducing spurious \(\sigma\) and redshift dependences.

In our formalism, \(f\) depends explicitly on both \(m\) and \(\sigma\); the artificial redshift dependence appears only when the mass dependence is forced to act exclusively through \(\sigma\). This leads to the \(\sim50\%\) variation shown in Fig.~\ref{hmf_sigma}, about four times larger than the typical modelling uncertainty of the (slightly simplified) framework used here. Since this behaviour follows directly from the explicit mass dependence predicted by the theory, we interpret the trend in Fig.~\ref{hmf_sigma} as lending strong support to the theoretical basis of our formalism.

\section{Performance of the theoretical framework and comparison with the Sheth $\&$ Tormen HMF}\label{results}

\begin{figure}
    \centering
    \includegraphics[width=1.0\linewidth]{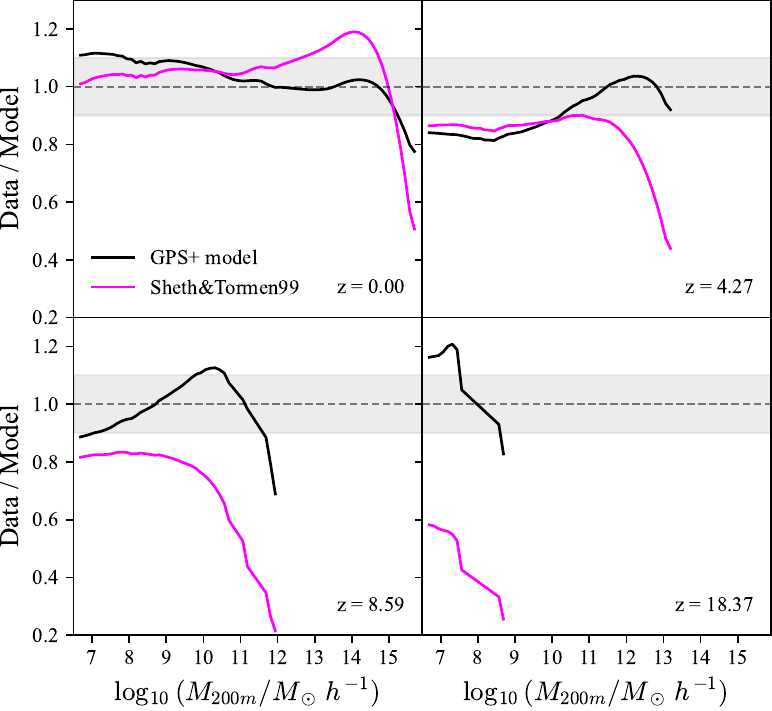}
   \caption{Ratios of the HMF predictions from GPS+ and ST at different redshifts. Only mass bins with $>100$ halos are shown.}

    \label{hmf_comparison}
\end{figure}

In this section, we present the HMF predicted by the theoretical framework developed and calibrated in this work, comparing it with the \textsc{Uchuu} simulation results described above; for reference, we also include the widely used Sheth–Tormen (ST) model\footnote{A comparison with the Dark emulator can be seen in Appendix \ref{dark_emulator_sec}.}, while further comparisons will be presented in Ishiyama et al. (in prep.). Figure~\ref{hmf_comparison} shows the ratio between the simulated HMF and the GPS+ prediction across a broad redshift range. At $z=0$, the ratio remains close to unity over most of the mass range, with deviations reaching $\sim20\%$ only at the high-mass end ($\log_{10}(M_{200m}/[h^{-1}M_\odot])>14$), while at $z=4.27$ similar deviations appear at the low-mass end ($\log_{10}(M_{200m}/[h^{-1}M_\odot])<11$). At higher redshifts, the agreement improves further, with deviations typically below $10\%$ across nearly the full mass range, demonstrating the robustness and predictive power of the GPS+ model over an exceptionally wide range in mass and redshift.

The ST model, which extends PS by incorporating ellipsoidal collapse and introduces an explicit redshift dependence through the linear growth factor (noting that ST adopts a friends-of-friends mass definition, whereas we use $M_{200m}$), performs similarly to GPS+ at $z=0$. At higher redshifts ($z \gtrsim 4.27$), however, GPS+ provides a markedly improved description, particularly for the most massive halos. This discrepancy increases toward earlier times: at $z \simeq 18.37$, ST deviates by $\sim60$–$80\%$, while GPS+ remains within $\sim10$–$20\%$. These results highlight both the accuracy of our framework across the full mass and redshift domains explored, along with the limitations of existing models when they are extrapolated to the high-redshift regime.

\section{Summary}\label{summary}
In this work, we present a physically motivated theoretical model for the halo mass function based on a generalised Press-Schechter framework with triaxial collapse (GPS+), which accurately reproduces the HMF measured in the \textsc{Uchuu} cosmological $N$-body simulation suite, spanning a wide range of volumes and mass resolutions within the \textit{Planck} cosmology. The model is rooted in the PS formalism, but it includes a modified collapse criterion for assigning mass elements to halos above a given mass, as detailed in the appendices. Unlike many previous approaches, GPS+ introduces no explicit redshift dependence, with evolution entering solely through the variance of the linear density field, $\sigma(M,z)$, enabling predictions up to $z\sim20$ over a halo mass range $6.5 \lesssim \log(M_{\rm h}) \lesssim 16$. Across this full mass and redshift domain, the model reproduces the simulated HMF with typical deviations below $10$–$20\%$, with the largest discrepancies confined to the high-mass end. Compared to the Sheth–Tormen model, GPS+ shows a similar accuracy at $z=0.0.$  However, it provides a substantially improved description at higher redshifts and for the most massive halos, where at $z=18.37$ Sheth–Tormen deviates by $70$-$80\%,$ while GPS+ remains within $20\%$. The uniqueness of this work lies both in the measurement of the HMF over an exceptionally wide mass and redshift range using the \textsc{Uchuu} simulations, paired with the ability of the GPS+ framework to predict the HMF across this range more reliably than any existing model.

We emphasise that our formalism is fully derived from first principles. Only $A$ and $B$ have been allowed to differ from their theoretical values, reflecting the details of our stabilisation model; thus, it is unsurprising that the best fit yields slightly different values.

\section{Data availability}
The GPS+ model is publicly available on the \textsc{Uchuu} GitHub\footnote{\url{https://github.com/uchuuproject/HMF_GPSplus}} and will be included in \textsc{Colossus}\footnote{\url{https://bdiemer.bitbucket.io/colossus/lss_mass_function.html}} \citep{2018ApJS..239...35D}. The combined HMF from the \textsc{Uchuu} suite is provided in electronic tables (Appendix~\ref{hmf_data}). The \textsc{Mucho-Uchuu-140M}, \textsc{1G}, and \textsc{6G} simulation products will be released on the \textsc{Skies \& Universes} website.

Table \ref{tab:HMF_summary} is only available in electronic form at the CDS via anonymous ftp to \url{cdsarc.u-strasbg.fr} (130.79.128.5) or via \url{http://cdsweb.u-strasbg.fr/cgi-bin/qcat?J/A+A/}.

\begin{acknowledgements}
This work was supported by project AST22\_00001\_27, funded by the European Union -- NextGenerationEU, the Spanish Ministry of Science, Innovation and Universities, the Recovery, Transformation and Resilience Plan, the Andalusian Regional Government (Junta de Andalucía), and the Spanish National Research Council (CSIC).

EFG acknowledges financial support from the Severo Ochoa programme (CEX2021-001131-S, MCIN/AEI/10.13039/501100011033). 

EFG and FP are supported by the Spanish MICINN grant PGC2018-101931-B-I00.

EFG and FP thank the Instituto de Astrofísica de Andalucía (IAA-CSIC), the Centro de Supercomputación de Galicia (CESGA), and RedIRIS for hosting the Uchuu DR1, DR2, and DR3 data releases on the \textsc{Skies \& Universes} platform.

TI acknowledges support from the IAAR Research Support Program at Chiba University, MEXT/JSPS KAKENHI (JP19KK0344, JP25H00662), the MEXT program ``Promoting Researches on the Supercomputer Fugaku'' (JPMXP1020230406), and JICFuS.

The Uchuu simulations were performed on the Aterui~II supercomputer at CfCA (NAOJ) and on the K computer at RIKEN, while the Mucho-Uchuu simulations were carried out on Fugaku at RIKEN (Project IDs: hp240184, hp250149). The Uchuu Data Releases made use of the IAA-CSIC computing facilities in Spain (MICINN EU-FEDER grant EQC2018-004366-P).
\end{acknowledgements}

%
%

   \bibliographystyle{aa} 
   \bibliography{example} 

@ARTICLE{1974ApJ...187..425P,
       author = {{Press}, William H. and {Schechter}, Paul},
        title = "{Formation of Galaxies and Clusters of Galaxies by Self-Similar Gravitational Condensation}",
      journal = {\apj},
         year = 1974,
        month = feb,
       volume = {187},
        pages = {425-438},
          doi = {10.1086/152650},
       adsurl = {https://ui.adsabs.harvard.edu/abs/1974ApJ...187..425P}
}

@ARTICLE{1999MNRAS.308..119S,
       author = {{Sheth}, Ravi K. and {Tormen}, Giuseppe},
        title = "{Large-scale bias and the peak background split}",
      journal = {\mnras},
         year = 1999,
        month = sep,
       volume = {308},
       number = {1},
        pages = {119-126},
          doi = {10.1046/j.1365-8711.1999.02692.x},
archivePrefix = {arXiv},
       eprint = {astro-ph/9901122},
 primaryClass = {astro-ph},
       adsurl = {https://ui.adsabs.harvard.edu/abs/1999MNRAS.308..119S}
}

@ARTICLE{2016MNRAS.463.1666C,
       author = {{Castro}, Tiago and {Marra}, Valerio and {Quartin}, Miguel},
        title = "{Constraining the halo mass function with observations}",
      journal = {\mnras},
         year = 2016,
        month = dec,
       volume = {463},
       number = {2},
        pages = {1666-1677},
          doi = {10.1093/mnras/stw2072},
archivePrefix = {arXiv},
       eprint = {1605.07548},
 primaryClass = {astro-ph.CO},
       adsurl = {https://ui.adsabs.harvard.edu/abs/2016MNRAS.463.1666C}
}

@ARTICLE{1998ApJ...495...80B,
       author = {{Bryan}, Greg L. and {Norman}, Michael L.},
        title = "{Statistical Properties of X-Ray Clusters: Analytic and Numerical Comparisons}",
      journal = {\apj},
         year = 1998,
        month = mar,
       volume = {495},
       number = {1},
        pages = {80-99},
          doi = {10.1086/305262},
archivePrefix = {arXiv},
       eprint = {astro-ph/9710107},
 primaryClass = {astro-ph},
       adsurl = {https://ui.adsabs.harvard.edu/abs/1998ApJ...495...80B}
}

@ARTICLE{2006ApJ...653L..77B,
       author = {{Betancort-Rijo}, Juan E. and {Montero-Dorta}, Antonio D.},
        title = "{The Halo Mass Function Redshift Dependence}",
      journal = {\apjl},
         year = 2006,
        month = dec,
       volume = {653},
       number = {2},
        pages = {L77-L80},
          doi = {10.1086/510582},
archivePrefix = {arXiv},
       eprint = {astro-ph/0608291},
 primaryClass = {astro-ph},
       adsurl = {https://ui.adsabs.harvard.edu/abs/2006ApJ...653L..77B}
}

@ARTICLE{2021MNRAS.506.4210I,
       author = {{Ishiyama}, Tomoaki and {Prada}, Francisco and {Klypin}, Anatoly A. and {Sinha}, Manodeep and {Metcalf}, R. Benton and {Jullo}, Eric and {Altieri}, Bruno and {Cora}, Sof{\'\i}a A. and {Croton}, Darren and {de la Torre}, Sylvain and {Mill{\'a}n-Calero}, David E. and {Oogi}, Taira and {Ruedas}, Jos{\'e} and {Vega-Mart{\'\i}nez}, Cristian A.},
        title = "{The Uchuu simulations: Data Release 1 and dark matter halo concentrations}",
      journal = {\mnras},
         year = 2021,
        month = sep,
       volume = {506},
       number = {3},
        pages = {4210-4231},
          doi = {10.1093/mnras/stab1755},
archivePrefix = {arXiv},
       eprint = {2007.14720},
 primaryClass = {astro-ph.CO},
       adsurl = {https://ui.adsabs.harvard.edu/abs/2021MNRAS.506.4210I}
}

@BOOK{2010gfe..book.....M,
       author = {{Mo}, Houjun and {van den Bosch}, Frank C. and {White}, Simon},
        title = "{Galaxy Formation and Evolution}",
         year = 2010,
          doi = {10.1017/CBO9780511807244},
       adsurl = {https://ui.adsabs.harvard.edu/abs/2010gfe..book.....M}
}

@ARTICLE{2002PhR...372....1C,
       author = {{Cooray}, Asantha and {Sheth}, Ravi},
        title = "{Halo models of large scale structure}",
      journal = {\physrep},
         year = 2002,
        month = dec,
       volume = {372},
       number = {1},
        pages = {1-129},
          doi = {10.1016/S0370-1573(02)00276-4},
archivePrefix = {arXiv},
       eprint = {astro-ph/0206508},
 primaryClass = {astro-ph},
       adsurl = {https://ui.adsabs.harvard.edu/abs/2002PhR...372....1C}
}

@ARTICLE{1991ApJ...379..440B,
       author = {{Bond}, J.~R. and {Cole}, S. and {Efstathiou}, G. and {Kaiser}, N.},
        title = "{Excursion Set Mass Functions for Hierarchical Gaussian Fluctuations}",
      journal = {\apj},
         year = 1991,
        month = oct,
       volume = {379},
        pages = {440},
          doi = {10.1086/170520},
       adsurl = {https://ui.adsabs.harvard.edu/abs/1991ApJ...379..440B}
}

@ARTICLE{1993MNRAS.262..627L,
       author = {{Lacey}, Cedric and {Cole}, Shaun},
        title = "{Merger rates in hierarchical models of galaxy formation}",
      journal = {\mnras},
         year = 1993,
        month = jun,
       volume = {262},
       number = {3},
        pages = {627-649},
          doi = {10.1093/mnras/262.3.627},
       adsurl = {https://ui.adsabs.harvard.edu/abs/1993MNRAS.262..627L}
}

@ARTICLE{2001MNRAS.323....1S,
       author = {{Sheth}, Ravi K. and {Mo}, H.~J. and {Tormen}, Giuseppe},
        title = "{Ellipsoidal collapse and an improved model for the number and spatial distribution of dark matter haloes}",
      journal = {\mnras},
         year = 2001,
        month = may,
       volume = {323},
       number = {1},
        pages = {1-12},
          doi = {10.1046/j.1365-8711.2001.04006.x},
archivePrefix = {arXiv},
       eprint = {astro-ph/9907024},
 primaryClass = {astro-ph},
       adsurl = {https://ui.adsabs.harvard.edu/abs/2001MNRAS.323....1S}
}

@ARTICLE{2001MNRAS.321..372J,
       author = {{Jenkins}, A. and {Frenk}, C.~S. and {White}, S.~D.~M. and {Colberg}, J.~M. and {Cole}, S. and {Evrard}, A.~E. and {Couchman}, H.~M.~P. and {Yoshida}, N.},
        title = "{The mass function of dark matter haloes}",
      journal = {\mnras},
         year = 2001,
        month = feb,
       volume = {321},
       number = {2},
        pages = {372-384},
          doi = {10.1046/j.1365-8711.2001.04029.x},
archivePrefix = {arXiv},
       eprint = {astro-ph/0005260},
 primaryClass = {astro-ph},
       adsurl = {https://ui.adsabs.harvard.edu/abs/2001MNRAS.321..372J}
}

@ARTICLE{2008ApJ...688..709T,
       author = {{Tinker}, Jeremy and {Kravtsov}, Andrey V. and {Klypin}, Anatoly and {Abazajian}, Kevork and {Warren}, Michael and {Yepes}, Gustavo and {Gottl{\"o}ber}, Stefan and {Holz}, Daniel E.},
        title = "{Toward a Halo Mass Function for Precision Cosmology: The Limits of Universality}",
      journal = {\apj},
         year = 2008,
        month = dec,
       volume = {688},
       number = {2},
        pages = {709-728},
          doi = {10.1086/591439},
archivePrefix = {arXiv},
       eprint = {0803.2706},
 primaryClass = {astro-ph},
       adsurl = {https://ui.adsabs.harvard.edu/abs/2008ApJ...688..709T}
}

@ARTICLE{2016MNRAS.456.2486D,
       author = {{Despali}, Giulia and {Giocoli}, Carlo and {Angulo}, Raul E. and {Tormen}, Giuseppe and {Sheth}, Ravi K. and {Baso}, Giacomo and {Moscardini}, Lauro},
        title = "{The universality of the virial halo mass function and models for non-universality of other halo definitions}",
      journal = {\mnras},
         year = 2016,
        month = mar,
       volume = {456},
       number = {3},
        pages = {2486-2504},
          doi = {10.1093/mnras/stv2842},
archivePrefix = {arXiv},
       eprint = {1507.05627},
 primaryClass = {astro-ph.CO},
       adsurl = {https://ui.adsabs.harvard.edu/abs/2016MNRAS.456.2486D}
}

@article{art3,
    author = {Ishiyama, Tomoaki and Fukushige, Toshiyuki and Makino, Junichiro},
    title = "{GreeM: Massively Parallel TreePM Code for Large Cosmological N-body Simulations}",
    journal = {Publications of the Astronomical Society of Japan},
    volume = {61},
    number = {6},
    pages = {1319-1330},
    year = {2009},
    month = {12},
    issn = {0004-6264},
    doi = {10.1093/pasj/61.6.1319},
}

@ARTICLE{art4,
       author = {{Ishiyama}, Tomoaki and {Nitadori}, Keigo and {Makino}, Junichiro},
        title = "{4.45 Pflops Astrophysical N-Body Simulation on K computer -- The Gravitational Trillion-Body Problem}",
      journal = {arXiv e-prints},
         year = 2012,
        month = nov,
          eid = {arXiv:1211.4406},
        pages = {arXiv:1211.4406},
          doi = {10.48550/arXiv.1211.4406},
archivePrefix = {arXiv},
       eprint = {1211.4406},
 primaryClass = {astro-ph.CO},
       adsurl = {https://ui.adsabs.harvard.edu/abs/2012arXiv1211.4406I}
}

@ARTICLE{Behroozi2013,
   author = {{Behroozi}, P.~S. and {Wechsler}, R.~H. and {Wu}, H.-Y.},
    title = "{The ROCKSTAR Phase-space Temporal Halo Finder and the Velocity Offsets of Cluster Cores}",
  journal = {\apj},
archivePrefix = "arXiv",
   eprint = {1110.4372},
 primaryClass = "astro-ph.CO",
     year = 2013,
    month = jan,
   volume = 762,
      eid = {109},
    pages = {109},
      doi = {10.1088/0004-637X/762/2/109},
   adsurl = {http://adsabs.harvard.edu/abs/2013ApJ...762..109B}
}

@ARTICLE{Behroozi2013b,
   author = {{Behroozi}, P.~S. and {Wechsler}, R.~H. and {Wu}, H.-Y. and 
{Busha}, M.~T. and {Klypin}, A.~A. and {Primack}, J.~R.},
    title = "{Gravitationally Consistent Halo Catalogs and Merger Trees for Precision Cosmology}",
  journal = {\apj},
archivePrefix = "arXiv",
   eprint = {1110.4370},
 primaryClass = "astro-ph.CO",
     year = 2013,
    month = jan,
   volume = 763,
      eid = {18},
    pages = {18},
      doi = {10.1088/0004-637X/763/1/18},
   adsurl = {http://adsabs.harvard.edu/abs/2013ApJ...763...18B}
}

@ARTICLE{2018ApJS..239...35D,
       author = {{Diemer}, Benedikt},
        title = "{COLOSSUS: A Python Toolkit for Cosmology, Large-scale Structure, and Dark Matter Halos}",
      journal = {\apjs},
         year = 2018,
        month = dec,
       volume = {239},
       number = {2},
          eid = {35},
        pages = {35},
          doi = {10.3847/1538-4365/aaee8c},
archivePrefix = {arXiv},
       eprint = {1712.04512},
 primaryClass = {astro-ph.CO},
       adsurl = {https://ui.adsabs.harvard.edu/abs/2018ApJS..239...35D}
}

@ARTICLE{2021A&A...652A.155S,
       author = {{Seppi}, R. and {Comparat}, J. and {Nandra}, K. and {Bulbul}, E. and {Prada}, F. and {Klypin}, A. and {Merloni}, A. and {Predehl}, P. and {Ider Chitham}, J.},
        title = "{The mass function dependence on the dynamical state of dark matter haloes}",
      journal = {\aap},
         year = 2021,
        month = aug,
       volume = {652},
          eid = {A155},
        pages = {A155},
          doi = {10.1051/0004-6361/202039123},
archivePrefix = {arXiv},
       eprint = {2008.03179},
 primaryClass = {astro-ph.CO},
       adsurl = {https://ui.adsabs.harvard.edu/abs/2021A&A...652A.155S}
}

@ARTICLE{2006ApJ...650L..95B,
       author = {{Betancort-Rijo}, Juan E. and {Montero-Dorta}, Antonio D.},
        title = "{Understanding the Cosmic Mass Function High-Mass Behavior}",
      journal = {\apjl},
         year = 2006,
        month = oct,
       volume = {650},
       number = {2},
        pages = {L95-L98},
          doi = {10.1086/507702},
archivePrefix = {arXiv},
       eprint = {astro-ph/0605608},
 primaryClass = {astro-ph},
       adsurl = {https://ui.adsabs.harvard.edu/abs/2006ApJ...650L..95B}
}

@ARTICLE{2000ApJ...534L.117B,
       author = {{Betancort-Rijo}, J. and {L{\'o}pez-Corredoira}, M.},
        title = "{The Complete Zeldovich Approximation}",
      journal = {\apjl},
         year = 2000,
        month = may,
       volume = {534},
       number = {2},
        pages = {L117-L121},
          doi = {10.1086/312679},
archivePrefix = {arXiv},
       eprint = {astro-ph/0003482},
 primaryClass = {astro-ph},
       adsurl = {https://ui.adsabs.harvard.edu/abs/2000ApJ...534L.117B}
}

@unpublished{MonteroDorta2006,
  author       = {Montero-Dorta, Antonio David and Betancort-Rijo, Juan},
  title        = {Advanced Studies Diploma by Antonio David Montero-Dorta, supervised by Juan Betancort-Rijo},
  year         = {2006},
  note         = {University of La Laguna, Tenerife, Spain},
  howpublished = {Diploma of Advanced Studies (DEA)}
}

@ARTICLE{2007MNRAS.378..339S,
       author = {{S{\'a}nchez-Conde}, M.~A. and {Betancort-Rijo}, J. and {Prada}, F.},
        title = "{The spherical collapse model with shell-crossing}",
      journal = {\mnras},
         year = 2007,
        month = jun,
       volume = {378},
       number = {1},
        pages = {339-352},
          doi = {10.1111/j.1365-2966.2007.11798.x},
archivePrefix = {arXiv},
       eprint = {astro-ph/0609479},
 primaryClass = {astro-ph},
       adsurl = {https://ui.adsabs.harvard.edu/abs/2007MNRAS.378..339S}
}

@ARTICLE{Shirasaki2021,
       author = {{Shirasaki}, Masato and {Ishiyama}, Tomoaki and {Ando}, Shin'ichiro},
        title = "{Virial Halo Mass Function in the Planck Cosmology}",
      journal = {\apj},
         year = 2021,
        month = nov,
       volume = {922},
       number = {1},
          eid = {89},
        pages = {89},
          doi = {10.3847/1538-4357/ac214b},
archivePrefix = {arXiv},
       eprint = {2108.11038},
 primaryClass = {astro-ph.CO},
       adsurl = {https://ui.adsabs.harvard.edu/abs/2021ApJ...922...89S}
}

@ARTICLE{Tokuue2024,
       author = {{Tokuue}, Tomoyuki and {Ishiyama}, Tomoaki and {Osato}, Ken and {Tanaka}, Satoshi and {Behroozi}, Peter},
        title = "{MPI-Rockstar: a Hybrid MPI and OpenMP Parallel Implementation of the Rockstar Halo finder}",
      journal = {arXiv e-prints},
         year = 2024,
        month = dec,
          eid = {arXiv:2412.18629},
        pages = {arXiv:2412.18629},
          doi = {10.48550/arXiv.2412.18629},
archivePrefix = {arXiv},
       eprint = {2412.18629},
 primaryClass = {astro-ph.IM},
       adsurl = {https://ui.adsabs.harvard.edu/abs/2024arXiv241218629T}
}

@ARTICLE{2020A&A...635A.135K,
       author = {{Kashibadze}, Olga G. and {Karachentsev}, Igor D. and {Karachentseva}, Valentina E.},
        title = "{Structure and kinematics of the Virgo cluster of galaxies}",
      journal = {\aap},
         year = 2020,
        month = mar,
       volume = {635},
          eid = {A135},
        pages = {A135},
          doi = {10.1051/0004-6361/201936172},
archivePrefix = {arXiv},
       eprint = {2002.12820},
 primaryClass = {astro-ph.GA},
       adsurl = {https://ui.adsabs.harvard.edu/abs/2020A&A...635A.135K}
}

@ARTICLE{2019ApJ...884...29N,
       author = {{Nishimichi}, Takahiro and {Takada}, Masahiro and {Takahashi}, Ryuichi and {Osato}, Ken and {Shirasaki}, Masato and {Oogi}, Taira and {Miyatake}, Hironao and {Oguri}, Masamune and {Murata}, Ryoma and {Kobayashi}, Yosuke and {Yoshida}, Naoki},
        title = "{Dark Quest. I. Fast and Accurate Emulation of Halo Clustering Statistics and Its Application to Galaxy Clustering}",
      journal = {\apj},
         year = 2019,
        month = oct,
       volume = {884},
       number = {1},
          eid = {29},
        pages = {29},
          doi = {10.3847/1538-4357/ab3719},
archivePrefix = {arXiv},
       eprint = {1811.09504},
 primaryClass = {astro-ph.CO},
       adsurl = {https://ui.adsabs.harvard.edu/abs/2019ApJ...884...29N}
}

\begin{appendix}

\section{Simulation Suite Summary}

Table \ref{sims} provides a summary of the cosmological $N$-body simulations used in this work. The simulation suite spans a wide range of volumes and mass resolutions, allowing us to probe dark matter halo statistics across several decades in mass and redshift while controlling for numerical and cosmological systematics.

The Phi-4096, Shin-Uchuu, and Uchuu simulations form a hierarchical set with increasing box size and decreasing mass resolution, all assuming a Planck15 cosmology. These simulations are particularly well suited for studying halo formation and clustering from galactic to large-scale structure scales within a consistent cosmological framework.

Complementing these runs, the Mucho-Uchuu simulations consist of multiple realizations at fixed volumes, adopting a Planck18 cosmology. The Mucho-Uchuu-140M, Mucho-Uchuu-1G, and Mucho-Uchuu-6G sets are designed to improve statistical precision through ensembles of realizations, enabling robust estimates of cosmic variance and covariance in halo statistics. Their differing box sizes and particle masses provide additional leverage to test resolution effects and volume dependence.

For each simulation, Table \ref{sims} lists the total number of particles, box size, dark matter particle mass, Plummer-equivalent gravitational softening length, number of realizations, adopted cosmological parameters, and relevant references. This diverse simulation suite allows for a comprehensive and systematic comparison between numerical results and theoretical predictions across a broad range of scales.

\begin{table*}[]
\caption{Summary of the simulation  properties.}
    \centering
    \begin{tabular}{cccccccc}
\hline\hline
        Name & N & L [Mpc$/h$] & M$_{\rm p}$ [M$_{\odot}/h$] & $\varepsilon$ $[h^{-1}$kpc$]$ & $\#$ & Cosmology & Ref.\\
        \midrule
         Phi-4096 & 4096$^{3}$ & 16 & 5.13 $\times$ 10$^{3}$ & 0.06 & 1 & Planck15 & \cite{2021MNRAS.506.4210I}\\
         Shin-Uchuu & 6400$^{3}$ & 140 & 8.96 $\times$ 10$^{5}$ & 0.4 & 1& Planck15 & \cite{2021MNRAS.506.4210I}\\
         Uchuu & 12800$^{3}$ & 2000 & 3.27 $\times$ 10$^{8}$ & 4.27 & 1 & Planck15 & \cite{2021MNRAS.506.4210I}\\
         Mucho-Uchuu-140M & 1024$^{3}$ & 140 & 2.21  $\times$  10$^{8}$ & 3.75 & 50 & Planck18 & Ishiyama et al., in prep.\\
         Mucho-Uchuu-1G & 4096$^{3}$ & 1000 & 1.26  $\times$  10$^{9}$ & 8 & 300 & Planck18 & Ishiyama et al., in prep.\\
         Mucho-Uchuu-6G & 6144$^{3}$ & 6000 & 8.04  $\times$  10$^{10}$ & 16 & 100 & Planck18 & Ishiyama et al., in prep.\\
\hline\hline
    \end{tabular}
    \tablefoot{Number of dark matter particles used in each simulation,  box size ($L_{\text{box}}$),  dark matter particle mass ($M_{\text{part}}$),  gravitational softening length ($\varepsilon$),  number of realisations ($\#$), and adopted cosmological parameters.}
    \label{sims}
\end{table*}

\section{Halo mass function from simulations}\label{hmf_data}
Table \ref{tab:HMF_summary} provides an example of the release HMF data, corresponding to redshift $z = 0$. This table, along with 19 additional ones (20 in total, with redshifts listed in Table A.2), is available in electronic form at the CDS. Each table includes the following columns: $M200m$, $f(M)$, $\Delta f(M)$, $\#$halos, Simulation, where $M200m$ is the  mass enclosed within overdensity 200$\rho_{\rm b}$; $f(M)$ is the multiplicity function (Equation \ref{multfunc}); $\Delta f(m)$ is its associated uncertainty; the number of halos corresponds to the count in each mass bin, $\#$; and the simulation name indicates which simulation contributed the HMF measurement for that bin.

\begin{table}[ht!]
\centering
\caption{HMF at $z=0.0$ from different cosmological simulations.}
\label{tab:HMF_summary}
\resizebox{\columnwidth}{!}{
\begin{tabular}{ccccc}
\hline\hline
$M_{200\rm b}$ [$M_\odot/h$] & $f(M)$ & $\Delta f(M)$ & \#halos & Simulation \\
\hline
4.9$\times10^{6}$ & 1.80$\times10^{-1}$ & 3.5$\times10^{-4}$ & 2.6$\times10^{5}$ & Phi-4096 \\
$\cdots$ & $\cdots$ & $\cdots$ & $\cdots$ & '' 
 \\
2.7$\times10^{8}$ & 2.23$\times10^{-1}$ & 2.7$\times10^{-3}$ & 7.1$\times10^{3}$ & Phi-4096 \\
\hline
3.7$\times10^{8}$ & 2.21$\times10^{-1}$ & 1.2$\times10^{-4}$ & 3.6$\times10^{6}$ & Shin-Uchuu \\
$\cdots$ & $\cdots$ & $\cdots$ & $\cdots$ & '' \\
8.7$\times10^{10}$ & 2.93$\times10^{-1}$ & 1.7$\times10^{-3}$ & 2.8$\times10^{4}$ & Shin-Uchuu \\
\hline
1.2$\times10^{11}$ & 2.98$\times10^{-1}$ & 2.8$\times10^{-3}$ & 4.8$\times10^{5}$ & MU-140M \\
$\cdots$ & $\cdots$ & $\cdots$ & $\cdots$ & '' \\
4.9$\times10^{11}$ & 3.27$\times10^{-1}$ & 5.4$\times10^{-3}$ & 1.4$\times10^{5}$ & MU-140M \\
\hline
6.5$\times10^{11}$ & 3.25$\times10^{-1}$ & 2.8$\times10^{-4}$ & 7.6$\times10^{7}$ & MU-1G \\
$\cdots$ & $\cdots$ & $\cdots$ & $\cdots$ & '' \\
2.7$\times10^{13}$ & 3.33$\times10^{-1}$ & 1.5$\times10^{-3}$ & 2.5$\times10^{6}$ & MU-1G \\
\hline
3.7$\times10^{13}$ & 3.23$\times10^{-1}$ & 1.4$\times10^{-4}$ & 4.0$\times10^{8}$ & MU-6G \\
$\cdots$ & $\cdots$ & $\cdots$ & $\cdots$ & '' \\
4.9$\times10^{15}$ & 1.07$\times10^{-4}$ & 1.7$\times10^{-5}$ & 1.6$\times10^{3}$ & MU-6G \\
\hline\hline
\end{tabular}
}

\tablefoot{Columns list: $M_{200m}$, the halo mass within overdensity $200\rho_{\rm b}$; 
$f(M)$, the mass function; 
$\Delta f(M)$, its uncertainty; the number of halos in each mass bin; and the name the name of the simulation contributing that bin. 
Only the first and last bins for each simulation are shown; ellipses indicate omitted intermediate values. 
Complete tables for all redshifts used in  Fig.~\ref{hmf_data_plot} are available in electronic form at the CDS.
}
\end{table}

\begin{table}[h!]
\caption{Available snapshots from each simulation.}
\centering
\tiny
\setlength{\tabcolsep}{3pt}
\renewcommand{\arraystretch}{0.85}
\begin{tabular}{@{}ccccccc@{}}
\toprule
$z$ & Phi-4096 & Shin-Uchuu & Uchuu & MU-140M & MU-1G & MU-6G \\ 
\midrule
0.00 & Y & Y & Y & Y & Y & Y \\
0.49 & Y & Y & Y & Y & - & - \\
1.03 & Y & Y & Y & Y & Y & Y \\
2.03 & Y & Y & Y & Y & Y & Y \\
2.95 & - & Y & Y & Y & Y & Y \\
4.27 & Y & Y & Y & Y & Y & - \\
5.15 & - & Y & Y & Y & Y & - \\
6.34 & - & Y & Y & Y & Y & - \\
7.03 & - & Y & Y & Y & Y & - \\
8.58 & Y & Y & Y & Y & Y & - \\
9.47 & Y & Y & Y & Y & Y & - \\
10.44 & Y & Y & Y & Y & Y & - \\
11.50 & Y & Y & Y & Y & Y & - \\
12.66 & Y & Y & Y & Y & Y & - \\
13.93 & Y & Y & Y & Y & - & - \\
14.32 & Y & Y & - & - & - & - \\
15.57 & Y & Y & - & - & - & - \\
16.92 & Y & Y & - & - & - & - \\
18.37 & Y & Y & - & - & - & - \\
19.95 & Y & Y & - & - & - & - \\
\bottomrule
\end{tabular}
\tablefoot{'Y' indicates that a snapshot exists at the corresponding redshift, and '-' indicates that no snapshot is available for that simulation. \textsc{MU} denotes the \textsc{Mucho-Uchuu} simulations.}
\label{redshifts}
\end{table}

\section{Basic considerations about the theoretical framework}\label{formalism_eqs}
The PS formalism yields a halo mass fraction $F(M)$ that is not correctly normalised ($F(M=0)=\frac{1}{2}$, instead of one). Furthermore, it predicts a high-mass tail that is too steep. This is rather puzzling because, while in other mass ranges there is ample room for improving the theoretical treatment, in this limit the spherical collapse approximation used in the PS formalism should become asymptotically exact. This is due to a selection effect: only proto-halos undergoing an almost perfectly spherical collapse have any chance of having formed by the time being considered.
ST combined ellipsoidal collapse with the PS approach and obtained a substantial improvement in the low-mass regime, but they did not solve either the high-mass problem or the normalisation issue.

In \cite{2006ApJ...650L..95B}, it was pointed out that the origin of these two problems lies in the PS assumptions themselves. The criterion used by those authors to decide whether a mass element belongs to an object with mass larger than $M$ was that a sphere with Lagrangian radius $r(M)$, centred on that element, satisfies the condition for spherical collapse (i.e. the linear fractional mass density fluctuation within the sphere, $\delta_{l}$, obeys $\geq\delta_{c}$). This criterion cannot be exact, since not all mass elements can be located at the centres of the objects that contain them. 

We adopted a more realistic and less restrictive criterion: a mass element belongs to a halo of mass $M$ if it lies within a distance $r(M)$ of a point that satisfies the collapse condition described above. This formulation allows an analytical treatment, leading to the mass fraction $F(m)$ expression in Equation~\ref{Fm}, where $V(\Sigma, m)/2$ represents the ratio between the probability that $\delta_{l}\geq1.686$ within a sphere with radius $r(M)$ centered on a randomly selected mass element and the greater probability that such an element belongs to an object with mass larger than $M$. In that work, we used the spherical collapse approximation, so that $<\delta_{c}>(\sigma, M)=\delta_{c}$ is constant (taken as either 1.686, or, 1.676). Under this formulation, both the high-mass limit and the normalisation behave correctly.

In \cite{MonteroDorta2006}, we used the Complete Zel'dovich Approximation (CZA) \citep{2000ApJ...534L.117B}, which effectively corresponds to ellipsoidal collapse, to compute the mean value of $\delta_{l}$ within the region (generally ellipsoidal) that collapses into an object of mass $M$. We denoted this mean value by $<\delta_{l}>$, and its variance by $U^{2}$. Both quantities were found to depend on $M$ and $z$ only through $\sigma(M, z)$, to a very good approximation). In the present work, the variance obtained in that study has been fitted using the expression given in Equation~\ref{Ux}. For $<\delta_{l}>$, the result obtained in that work was

\begin{equation}
    <\delta_{l}> = \delta_{c}\left[0.814+0.688\left(\frac{\sigma}{\delta_{c}}\right) -0.033\left(\frac{\sigma}{\delta_{c}}\right)^{2}+0.02\left(\frac{\sigma}{\delta_{c}}\right)^{3}   \right]
    \label{B1}
.\end{equation}

This expression corresponds to the first factor in Equation~\ref{delta_c}. The second factor introduces  a small explicit dependence on $M$ whose origin and derivation will be presented in Betancort-Rijo et al. in prep..

Equation \ref{B1} was obtained using a model for shell-crossing and stabilisation in which the protohalo, upon collapsing along one direction in comoving coordinates to $\frac{1}{13}$ of its original size, becomes stabilised along that axis while continuing to collapse along the remaining directions.
This stabilisation model  is physically well motivated but not exact, so we  introduced the parameters $A$ and $B$, which are expected to remain close to unity, to account for its approximations. 

It is important to note that in this formalism $F(M)$ depends on mass not only through $\sigma(M, z)$, but also through an explicit dependence on $M$ that reflects the shape of the power spectrum at that scale $M$. This dependence appears in Equation \ref{delta_c} through $b(M)$ (\ref{bm}) and in $V(M)$ (Equation \ref{Vm}) through $c(M)$ (Equation \ref{cm}). 

It should also be emphasised that there is no explicit dependence on redshift, and that $F(M)$ is correctly normalised without requiring parameters fitting. The only fitted parameters are $A$ and $B$ in $<\delta_{c}>(\sigma, M)$, which, based on theoretical considerations, are expected to remain close to unity. In principle, their values could be determined directly from simulations by analysing the behaviour of this function, since analytically we cannot extend the derivation beyond \ref{B1}.

\section{Multiplicity function as a function of $\sigma(M)$}
\begin{figure}[b!]
    \centering
    \includegraphics[width=1.0\linewidth]{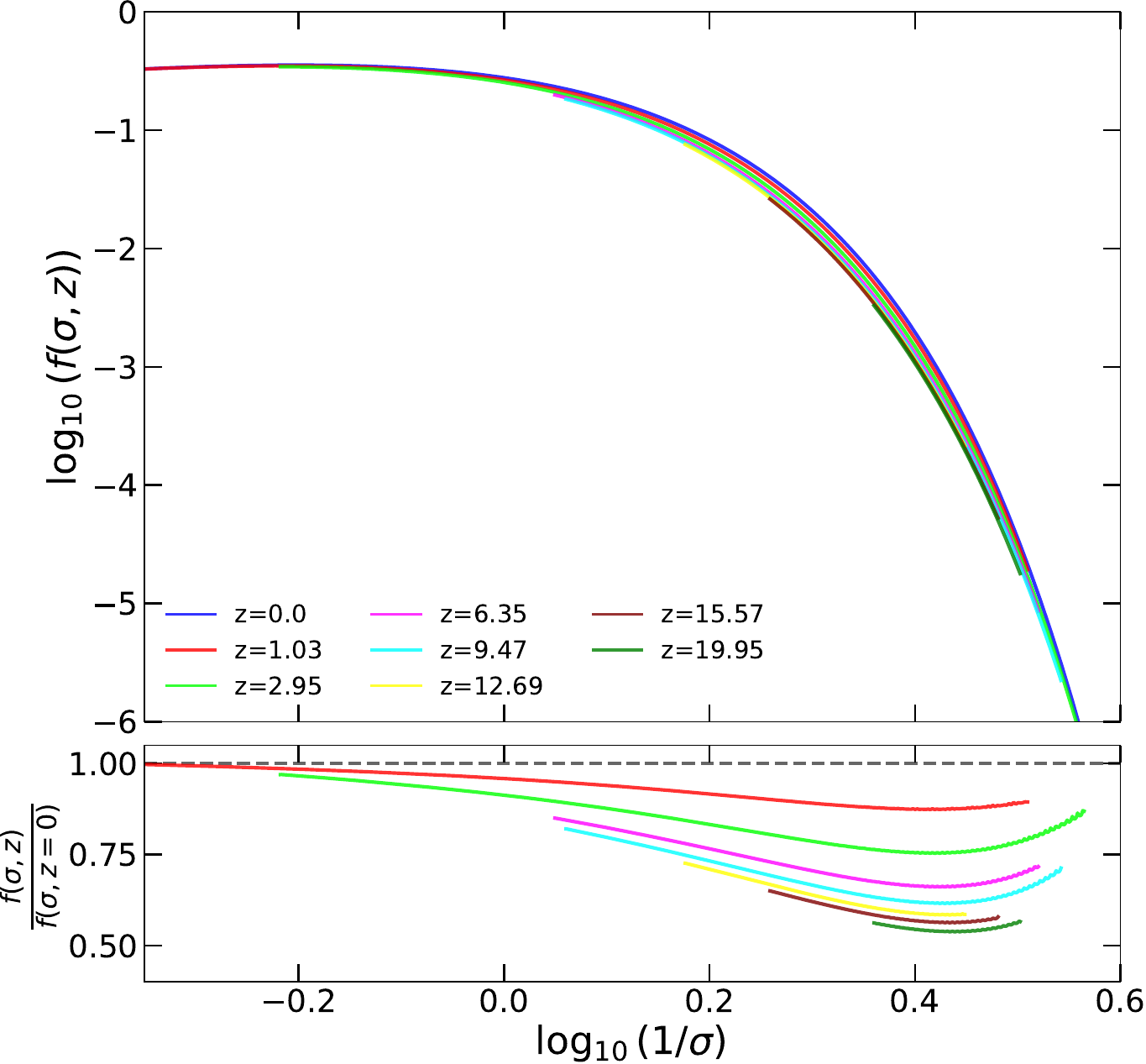}
    \caption{Top panel: Multiplicity function $f(\sigma, z)$ as a function of $\sigma$ at different redshifts, as predicted by the GPS+ model presented in this work. Bottom panel:\ Ratio of the model at each redshift to $z=0$ prediction.}
    \label{hmf_sigma}
\end{figure}

In this appendix, we show the multiplicity function, $f(\sigma, z),$ as a function of $\sigma$ at different redshifts, as predicted by the GPS+ model presented in this work (Figure \ref{hmf_sigma}).

\section{Dependence of the HMF with the halo mass definition}\label{mass_def}
\begin{figure}
    \centering
    \includegraphics[width=1.0\linewidth]{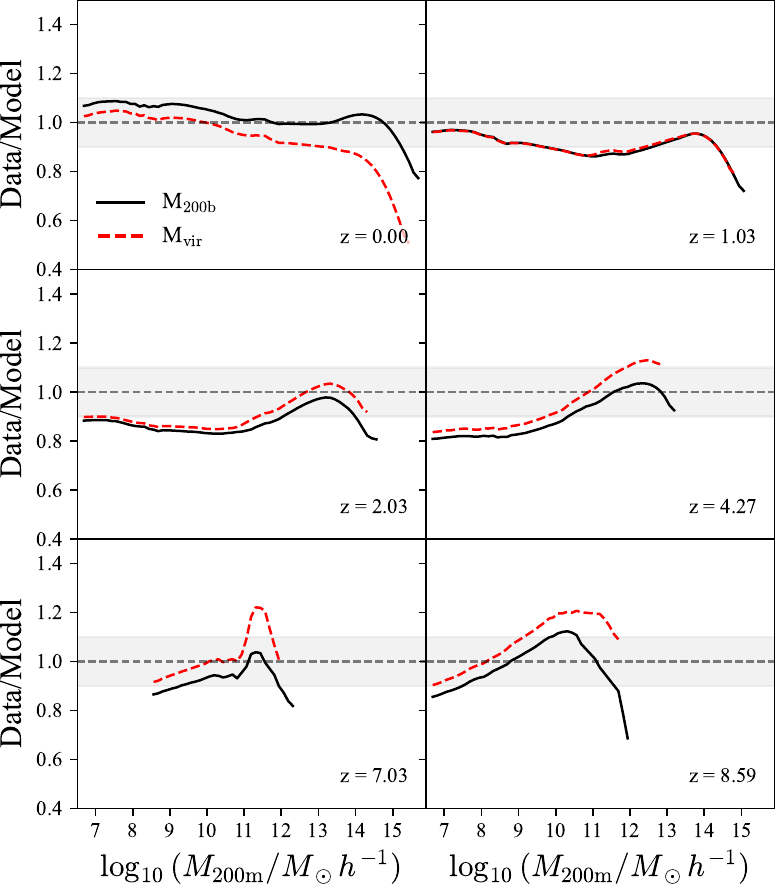}
    \caption{Comparison of the ratio between the simulated HMF and the GPS+ model at different redshifts, for two halo mass definitions in the data: the mass enclosed within overdensity 200$\rho_{\rm b}$, M$_{\rm 200m}$, and the mass within the virial radius, M$_{\rm vir}$.}
    \label{mass_defs_comp}
\end{figure}

As part of this work, we assessed the impact of different halo mass definitions in our theoretical framework. The HMF results presented in Section \ref{results} are based on the M$_{\rm 200m}$ mass definition; however, the virial mass, M$_{\rm vir}$, can also be adopted. Figure \ref{mass_defs_comp} compares the performance of the model under both definitions. The largest differences arise at low redshift. When using the virial mass, the agreement at the high-mass end worsens significantly, with ratios dropping to $\sim0.5$, corresponding to discrepancies of $\sim50\%$ between the model predictions and the measured HMFs from the simulations. At higher redshifts, the performance of the model using the virial mass becomes comparable to that obtained using the M$_{\rm 200m}$  definition. 

This analysis shows that the choice of halo mass definition has a strong impact on the performance of the model, particulary at low redshifts, where the difference between M$_{\rm vir}$ and M$_{\rm 200m}$ becomes significant. The virial mass definition does not appear to be suitable in this context, as no choise of fitted parameters yields a good agreement at the high-mass end of the HMF at $z=0.0$. This behaviour is related to the fact that the virial overdensity, $\Delta_{\rm vir}$, is not constant. In the \textsc{Uchuu} simulation suite, $\Delta_{\rm vir}$ evolves following \cite{1998ApJ...495...80B}, giving $\Delta_{\rm vir}(z=0)\approx102$, $\Delta_{\rm vir}(z=1.03)\approx159$, and approaching the Einstein–de Sitter value of 178 at higher redshift. That work argues that the density contrast at virialisation should depend on $\Omega_{\rm m}$ (and thus on redshift for $\Lambda$CDM cosmologies). However, there are strong counterarguments suggesting that $\Delta_{\rm vir}$ should be nearly independent of redshift. A detailed discussion of this point and its theoretical implications will be presented in Betancort-Rijo et al. in prep.; here, we simply highlight that the available evidence points toward a weak or negligible redshift dependence of $\Delta_{\rm vir}$. In \cite{2007MNRAS.378..339S}, we investigated the spherical collapse model with shell-crossing to study the stabilisation of proto-halos and no difference  was found amongst the Einstein-de Sitter and $\Lambda$CDM cosmologies.

If the results of \cite{1998ApJ...495...80B} were correct and the stabilisation process truly depended on redshift, then the HMF constructed using M$_{\rm 200m}$ should exhibit explicit redshift dependence. In that case, the formalism presented here - which contains no explicit dependece on $z$ - would be unable to reproduce the simulation results accurately. Conversely, the HMF constructed using M$_{\rm vir}$, together with its redshift-dependent $\Delta_{\rm vir}$, would account for the assumed redshift dependence of the stabilisation process and thus eliminate this source of spurious explicit redshift dependence. Under such circumstances, the HMF based on M$_{\rm vir}$ would be expected to appear more universal.

However, Figure~\ref{mass_defs_comp} shows the opposite behavior: the HMF constructed using M$_{\rm 200m}$ provides  a significantly better match to the formalism (which is explicitly independent of $z$), and therefore yields a more universal HMF. 

It should be noted that in Figure \ref{mass_defs_comp}, the M$_{\rm vir}$ data from the simulations are compared with the predictions of the formalism using parameter values slightly adjusted relative to those employed for M$_{\rm 200m}$, in order to obtain the best possible agreement across the full redshift range. Even with this optimisation, the model fails to accurately reproduce the HMF at low redshift when using M$_{\rm vir}$ at $z=0$, there is a pronounced discrepancy for high-mass end, while for $z=1.03$ and larger z's the difference residues for both mass definitions are within the accuracy of the formalism. Since for these redshifts $\Delta_{\rm vir}$ changes little according to \cite{1998ApJ...495...80B} formalism, this support the conclusion that $\Delta_{\rm vir}$ is in fact almost independent of redshift.

\section{Comparison with the Dark Emulator}\label{dark_emulator_sec}

\begin{figure}
    \centering
    \includegraphics[width=\linewidth]{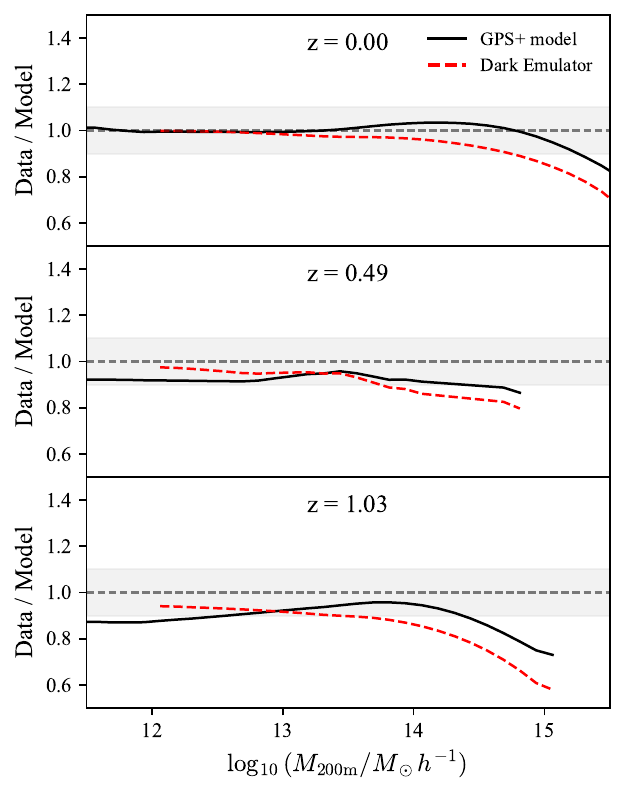}
    \caption{Comparison of halo mass function ratios using two different theoretical predictions. Each panel shows the ratio of the data to the GPS+model prediction and to the Dark Emulator prediction, highlighting relative differences between the two approaches. The panels correspond to redshifts $z = 0.0$ (left), $z = 0.49$ (middle), and $z = 1.03$ (right).}
    \label{darkemulator}
\end{figure}

We compare the GPS+ model with the predictions from the widely used Dark Emulator \citep{2019ApJ...884...29N}, which is calibrated only up to $z = 1.48$ and for M$_{\rm 200m}\gtrsim10^{12} h^{-1}M_{\odot}$; therefore, the comparison is restricted to redshifts below this limit.

Figure~\ref{darkemulator} presents the ratios of the data relative to the HMF predicted by the GPS+ model and by the Dark Emulator at $z = 0.0$ (left panel), $z = 0.49$ (middle panel), and $z = 1.03$ (right panel). The Dark Emulator fails to reproduce the \textsc{Uchuu} HMF at low halo masses and systematically overpredicts the HMF at the high-mass end. Overall, the GPS+ model shows significantly better agreement with the simulation data across the full mass range considered.

\end{appendix}

\end{document}